\documentclass[article]{aa501}
\usepackage{amsmath,amssymb}
\usepackage{graphicx}
\usepackage{natbib}

\newcommand{\rmd}{{\rm d}}

\newcommand{\bx}{{\mathbf{x}}}

\newcommand{\BR}{{\mathbb{R}}}

\newcommand{\CM}{{\cal M}}

\newcommand{\hMpc}{{\ifmmode{h^{-1}{\rm Mpc}}\else{$h^{-1}$Mpc}\fi}}

\newcommand{\paverage}[1]{\left\langle #1 \right\rangle_{\rm P}}
\newcommand{\cov}{{\ifmmode{\text{{\it cov}}}\else{ {\it cov} }\fi}}
\newcommand{\var}{{\ifmmode{\text{{\it var}}}\else{ {\it var} }\fi}}

\begin{document}
\title{Spatial distribution of galactic halos 
and their merger histories}

\author{
S.~Gottl\"ober\inst{1} \and
M.~Kerscher\inst{2} \and 
A.V.~Kravtsov\inst{3,4} \and
A.~Faltenbacher\inst{1}  \and
A.~Klypin \inst{5} \and
V.~M\"uller\inst{1} 
}

\offprints{S. Gottl\"ober}

\institute{
Astrophysikalisches Institut Potsdam, 
An der Sternwarte 16, 14482 Potsdam,Germany
\and
Sektion Physik, Ludwig--Maximilians--Universit{\"a}t, 
Theresienstra{\ss}e 37, D--80333 M{\"u}nchen, Germany
\and
Department of Astronomy \& Astrophysics,
The University of Chicago, 5640 S. Ellis Ave. 
Chicago, IL 60637, USA
\and
 Center for Cosmological Physics, Enrico Fermi Institute, 
The University of Chicago, IL 60637
\and
Astronomy Department, New Mexico State University, Las Cruces, NM
88003-0001, USA
}

\date{Received February 22, 2002; accepted March 05, 2002}


\abstract{We use a novel statistical tool, the mark correlation
  functions (MCFs), to study clustering of galaxy-size halos as a
  function of their properties and environment in a high-resolution
  numerical simulation of the $\Lambda$CDM cosmology. We applied MCFs
  using several types of continuous and discrete marks: maximum
  circular velocity of halos, merger mark indicating whether halos
  experienced or not a major merger in their evolution history (the
  marks for halo with mergers are further split according to the epoch
  of the last major merger), and a stripping mark indicating whether
  the halo underwent a tidal stripping (i.e., mass loss). We find
  that halos which experienced a relatively early ($z>1$) major merger
  or mass loss (due to tidal stripping) in their evolution histories
  are over-abundant in halo pairs with separations $\lesssim 3${\hMpc}. 
  This result can be interpreted as spatial segregation of halos with
  different merger histories, qualitatively similar to the
  morphological segregation in the observed galaxy distribution.  In
  addition, we find that at $z=0$ the mean circular velocity of halos
  in pairs of halos with separations $\lesssim 10$\hMpc\ is larger
  than the mean circular velocity $\overline{v}_{\rm circ}$ of the
  parent halo sample.  This mean circular velocity enhancement
  increases steadily during the evolution of halos from $z=3$ to
  $z=0$, and indicates that the luminosity dependence of galaxy
  clustering may be due to the mass segregation of galactic dark
  matter halos. The analysis presented in this paper demonstrate that MCFs
  provide powerful, yet algorithmically simple, quantitative measures
  of segregation in the spatial distribution of objects with respect to 
  their various properties (marks). This should make MCFs very useful
  for analysis of spatial clustering and segregation in current and
  future large redshift surveys.
}

\maketitle

\section{Introduction}

The advent of large wide-field redshift surveys of galaxies, such as
the Two-Degree Field \citep[2dF,][]{colless_etal01} and the Sloan
Digital Sky Survey \citep[SDSS,][]{york:sloan}, will allow detailed
studies of clustering of galaxies as a function of their environment
and internal properties.  Indeed, hierarchical growth of structure via
gravitational instability is thought to play a dominant role in shaping
both the large-scale galaxy clustering and internal properties of
galaxies such as luminosity, colors, and morphology. This close
connection implies that studies of the spatial distribution of galaxies
as a function of their internal properties and environment should
provide us valuable insights into the process of galaxy formation.
Previous observational studies and the first results from the 2dF and
SDSS have shown that clustering strength depends on morphology
\citep[e.g.,][]{hermit_etal96,guzzo_etal97}, luminosity (see, e.g.,
{}\citealt{hamilton:evidence,benoist:biasing,norberg_etal01}), and
colors \citep[e.g.,][]{zehavi_etal02} of galaxies.  Greatly enhanced
clustering of super-luminous IR-galaxies {}\citep{bouchet:slir} and
morphological
\citep{dressler:galaxy,postman_geller84,whitmore_etal93,biviano:theeso}
and color
\citep[e.g.,][]{butcher:evolution,margonier:butcher} segregation in
clusters of galaxies indicate dependence of clustering on the merging
history and large-scale environment.

In this paper we use a novel statistical tool, the mark correlation
functions (hereafter MCFs), to study clustering of galactic halos as a
function of their properties and environment in a high-resolution
numerical simulation of the $\Lambda$CDM cosmology.  Mark correlation
functions \citep{stoyan:oncorrelations,stoyan:fractals} have been
introduced into astrophysics only recently
\citep{beisbart:luminosity}, although some aspects of marked point
processes were discussed by \citet{peebles:lss}. The mark statistics can
be used to quantify the differences in the spatial distributions of
various galaxy samples (similarly to the usual two-point correlation
function) and, at the same time, to study the interplay between the spatial
clustering and the distribution of galaxy properties (marks).  In this
respect, the MCFs are the natural extension of
the spatial correlation functions for studies where it is advantageous
to consider mark and spatial distributions simultaneously.  Variants
of the MCFs can be used to study both continuous (e.g., luminosity or
angular momentum) and discrete distributions (e.g., color classes or
morphological types) of galactic properties. This makes them valuable
for quantitative studies of luminosity and morphological segregation
of galaxies as well as dependence of spatial distribution of galaxies
on various events in their evolutionary history (e.g., time since the
last major merger), which can be used as marks.
Indeed, the mark correlation statistics quickly proved to be a very
useful tool for identification of physical processes that shape the
observed spatial distribution of galaxies
({}\citealt{szapudi:correlationspscz}, see
{}\citealt{beisbart:wuppertal} for a recent review). 

The first step towards the use of clustering as a probe of processes
shaping properties of galaxies is a good theoretical understanding of
how these processes affect spatial distribution of galaxies. During the
last several years, thanks to continuously improving spatial and mass
resolutions of numerical simulations and development of semi-analytic
models of galaxy formation, there was a significant progress in the
theoretical understanding of galaxy clustering evolution and bias
\citep[e.g.,][]{bagla:evolution,jing98,kauffmann:clusteringI,katz:clustering,colin:evolution,kravtsov:origin,pearce:simulation,schmalzing:quantifying}.
The complete information about the internal properties and evolution
of galactic halos, usually available in theoretical analysis, allows
one to study in detail the interplay between different evolutionary
processes and spatial distribution of objects. In the present paper we
use mark correlation functions to study clustering of galaxy-size dark
matter halos and its dependence on the halo properties (e.g., circular
velocity) and evolution history in a high-resolution simulation of the
currently favored flat Cold Dark Matter (CDM) model with cosmological
constant (see \S~\ref{sect:numsim}).

The paper is organized as follows. We discuss briefly the numerical
simulation in Sect.~\ref{sect:numsim}. In
Sect.~\ref{sect:markcorr} we introduce and explain the properties of
mark correlation functions. In Sect.~\ref{sect:results} we present
analysis of the spatial distribution of dark matter halos using mark
correlation functions. In Sect.~\ref{sect:conclusions} we discuss
results and their implications and summarize our conclusions.

\section{Numerical simulations}
\label{sect:numsim}

We used the Adaptive Refinement Tree (ART) code
\citep{kravtsov:adaptive} to simulate the evolution of collisionless DM
in the currently favored $\Lambda$CDM model ($\Omega_{\rm
  m}=1-\Omega_{\Lambda}=0.3$; $H_0=h \times 100$km s$^{-1}$ Mpc$^{-1}
=70$km s$^{-1}$ Mpc$^{-1}$;
$\sigma_8=1.0$). The age of the Universe in this cosmology is 13.5
Gyrs and normalization is in accordance with the four year {\sl COBE}
DMR observations \citep{bunn:fouryear} as well as the observed
abundance of galaxy clusters
\citep[e.g.,][]{pierpaoli:power,ikebe:new}.  

The numerical simulation of the $\Lambda$CDM model followed the
evolution of $256^3\approx 1.67\times 10^7$ particles in a periodic
$60\hMpc$ box.  The particle mass is thus $\approx 1.1\times
10^9h^{-1}{\rm M_{\odot}}$.  The ART code reaches high force
resolution by refining all high-density regions with an automated
refinement algorithm.  The refinements are recursive: the refined
regions can also be refined, each subsequent refinement having half of
the previous level's cell size.  This creates an hierarchy of
refinement meshes of different resolution covering regions of
interest.  The criterion for refinement is {\em local overdensity} of
particles: in the simulation presented in this paper the code refined
an individual cell only if the density of particles (smoothed with the
cloud-in-cell scheme) was higher than $n_{th}=5$ particles. Therefore,
{\em all} regions with overdensity higher than $\delta = n_{th}{\ 
  }2^{3L}/\bar{n}$, where $\bar{n}$ is the average number density of
particles in the cube, were refined to the refinement level $L$. For
the simulation presented here, $\bar{n}$ is $1/8$.  The peak formal
dynamic range reached by the code on the highest refinement level is
$32,768$, which corresponds to the smallest grid cell of
$1.83h^{-1}{\rm\ kpc}$; the actual force resolution is approximately a
factor of two lower. The simulation that we
analyze here has been used in \citet{colin:evolution}, and we refer
the reader to this paper for further numerical details.

Identification of DM halos in the very high-density environments (e.g.,
inside groups and clusters) is a challenging problem. The goal of this
study is to investigate spatial correlations of halo populations as
closely related to the observed galaxy population as possible.  This
requires identification of both isolated halos and satellite halos
orbiting within the virial radii of larger systems.  The problems
associated with halo identification within high-density regions are
discussed in \citet{klypin:galaxies}. In this study we use a halo
finding algorithm called Bound Density Maxima (BDM). The main idea of
the BDM algorithm is to find positions of local maxima in the density
field smoothed at a certain scale and to apply physically motivated
criteria to test whether the identified site corresponds to a
gravitationally bound halo\footnote{The detailed description of the
algorithm is given in \citep{klypin:galaxies} and
\citep{colin:evolution}.}.  It is based on the ideas of the DENMAX halo
finder {}\citep{bertschinger:denmax}, in the sense that the BDM makes
sure that the density peaks are gravitationally bound and estimates
parameters of the halos after removing unbound particles.  The
algorithm identifies both isolated halos and subhalos located in the
virial regions of more massive halos.  The distribution of halos
identified in this way can be compared to the distribution of galaxies
directly, because the halo and galaxy catalogs include both isolated
systems and objects within clusters and groups.

Even with algorithms tailored for identification of sub-halos,
additional problems exist. Interacting halos exchange and loose mass;
the total mass of a halo depends on its radius, which is difficult to
define in a dense environment within virialized regions.  We alleviate
the latter problem by using the maximum circular velocity instead
of the mass. In practice, maximum circular velocity is a rather stable
quantity which changes little even when halos looses most of its mass
and can serve therefore as a useful mass-related ``tag'' of a halo.
Numerically, the maximum ``circular velocity'' ($\sqrt{GM/R}$), $v_{\rm
circ}$ can be measured more accurately then mass. In addition, the
maximum circular velocity can be more readily compared to observations
than, for example, virial mass or mass within the tidal radius of the
halo.

\section{Mark correlation functions}
\label{sect:markcorr}

In studying galaxy clustering with the mark correlation functions, we
view galaxies as discrete points in space with marks describing their
intrinsic physical properties. Thus, we consider a point set
$\{\bx_i\}_{i=1}^{N}$ and attach a mark $m_i$ to each point
$\bx_i\in\BR^3$ resulting in the marked point set
$\{(\bx_i,m_i)\}_{i=1}^{N}$
{}\citep{stoyan:oncorrelations,stoyan:fractals}. The marks, in turn,
can be either continuous or discrete.  In the following, we use the
circular velocity as a continuous mark and merging/stripping events of
halos as discrete marks. In a subsequent paper, we will apply MCFs 
to study various other marks, such as
the spin parameter (continuous 
scalar mark) and the angular momentum (vector mark).

Let $\varrho$ be the mean number density of the points in space and
$\varrho^M(m)\rmd{}m$ the probability that the value of a mark on a
point lies within the interval $[m,m+\rmd m]$. The mean mark is then
$\overline{m}=\int\rmd m\ \varrho^M(m)m$, the mark variance is
$\sigma_M^2=\int\rmd m\ \varrho^M(m) (m-\overline{m})^2$.
We assume that the joint probability $\varrho^{SM}(\bx,m)$ of finding a
point   at   position  $\bx$   with   mark   $M$,   splits  into   a
space--independent  mark probability  and the mean density:
$\varrho^M(m)\times\varrho$.
The spatial--mark product--density
\begin{equation}
\varrho_2^{SM}((\bx_1,m_1),(\bx_2,m_2))\ 
\rmd V_1 \rmd m_1\ \rmd V_2 \rmd m_2 ,
\end{equation}
is the joint probability of finding  a point at $\bx_1$ with the mark
$m_1$ and another point at $\bx_2$ with the mark $m_2$.  We obtain
the spatial  product density $\varrho_2(\bx_1,\bx_2)$  and the two--point
correlation function $\xi(r)$ by marginalizing over the marks:
\begin{multline}
\varrho^2\ (1+\xi(r)) = \varrho_2(\bx_1,\bx_2) = \\
=\int\rmd m_1 \int\rmd m_2\ \varrho_2^{SM}((\bx_1,m_1),(\bx_2,m_2)),
\end{multline}
where $\xi(r)$ is the spatial two-point correlation function 
which depends only on the separation $r=|\bx_1-\bx_2|$ of the
points for a homogeneous and isotropic point set.

We define the conditional mark density:
\begin{multline}
\CM_2(m_1,m_2|\bx_1,\bx_2) =\\
=\begin{cases}
\frac{\varrho_2^{SM}((\bx_1,m_1),(\bx_2,m_2))}{\varrho_2(\bx_1,\bx_2)}
& \text{ for } \varrho_2(\bx_1,\bx_2)\ne 0,\\
0 & \text{ else }.
\end{cases} 
\end{multline}
For     a    stationary     and    isotropic     point    distribution
$\CM_2(m_1,m_2|r)\rmd m_1\rmd m_2$ is the probability of finding
the marks $m_1$  and $m_2$ at two points located at $\bx_1$ and
$\bx_2$,   under   the   condition   that  they   are   separated   by
$r=|\bx_1-\bx_2|$.
Now the full mark product--density can be written as
\begin{equation}
\varrho_2^{SM}((\bx_1,m_1),(\bx_2,m_2)) 
= \CM_2(m_1,m_2|\bx_1,\bx_2)\ \varrho_2(\bx_1,\bx_2) .
\end{equation}
If there is no mark segregation in space, $\CM_2(m_1,m_2|r)$ is independent
of $r$, and $\CM_2(m_1,m_2|r)=\varrho^M(m_1)\varrho^M(m_2)$.

Starting from these definitions, especially using the conditional mark
density $\CM_2(m_1,m_2|r)$, one may construct several
mark correlation functions sensitive to different aspects of
mark segregation {}\citep{beisbart:luminosity}.  The basic idea is
to consider weighted conditional correlation functions describing
the probability of finding two points at a separation $r$.
For a positively defined weighting  function $f(m_1,m_2)\geq 0$ we define the
average over pairs with separation $r$:
\begin{equation}
\label{eq:def-paverage}
\paverage{f}(r) = \int\rmd m_1\int\rmd m_2\ f(m_1,m_2)\ 
\CM_2(m_1,m_2|r).
\end{equation}
$\paverage{f}(r)$ is the expectation value of the weighting function
$f$ (depending only on the marks), under the condition that we find a
point pair with separation $r$.  For a suitable defined
integration measure Eq.~\eqref{eq:def-paverage} is also applicable to
discrete marks.
The definition~\eqref{eq:def-paverage} is very flexible, and allows us
to investigate the correlations of both continuous and discrete 
marks.

In the following analysis, we calculated the mark correlation functions
taking into account the periodicity of the simulation box. However, we
obtain virtually identical results using the estimator without boundary
corrections (see Appendix A of {}\citet{beisbart:luminosity} for
details).

\subsection{Correlations of scalar marks}

For scalar marks the following mark correlation functions have proven
to be useful
({}\citealt{stoyan:fractals}, {}\citealt{beisbart:luminosity}, 
{}\citealt{schlather:mark}):
The simplest weight to be used is the mean mark:
\begin{equation}
k_{m}(r) \equiv \frac{\paverage{m_1+m_2}(r)}{2\ \overline{m}} .
\label{eq:k_m}
\end{equation}
It quantifies the deviation of the mean mark on pairs with separation
$r$ from the overall mean mark $\overline{m}$.  For example, $k_{m}(r)>1$
indicates mark segregation for point pairs with a separation $r$,
specifically their mean mark is larger than the overall mark average.\\
Higher moments of marks like the mark fluctuations
\begin{equation}
\var(r) \equiv \paverage{\left(m_1-\paverage{m_1}(r)\right)^2}(r) ,
\label{eq:var}
\end{equation}
or the mark covariance {}\citep{cressie:statistics} 
\begin{align}
\label{eq:def-cov}
\cov(r) & \equiv \paverage{ \big(m_1-\paverage{m_1}(r)\big)\big(m_2-\paverage{m_2}(r)\big)}(r) \nonumber \\
        & =  \paverage{m_1 m_2}(r) - \paverage{m_1}(r)\paverage{m_2}(r),
\end{align}
may be used to quantify mark segregation. For example, a positive
$\cov(r)$ indicates that points with separation $r$ tend to have
similar marks, whereas a negative $\cov(r)$ indicates different marks.

\subsection{Correlations of discrete marks}

For discrete labels only combinations of indicator--functions are
possible, and the integral degenerates into a sum over the labels.
Supposing the marks of our objects belong to classes labeled with
$i,j,\ldots$, the conditional cross--correlation functions are given by
\begin{equation}
\label{eq:cond-crosscorr}
C_{i j}(r) \equiv
\paverage{\delta_{m_1i} \delta_{m_2j} +
(1-\delta_{ij}) \delta_{m_2i} \delta_{m_1j}} (r),
\end{equation}
with  the Kronecker $\delta_{m_1i}=1$  for $m_1=i$  and zero
otherwise {}\citep{stoyan:fractals}. By construction
\begin{equation}
\sum_i\sum_j C_{i,j}(r) = 1 .
\end{equation}
Mark segregation is indicated by
$C_{ij}\ne2\varrho_i\varrho_j/\varrho^2$ for $i\ne j$ and
$C_{ii}\ne\varrho_i^2/\varrho^2$, where $\varrho_i$ denotes the number
density of points with label $i$.  The $C_{ij}$ are cross--correlation
functions under the {\em condition} that two points are separated by a
distance of~$r$.

\section{Results}
\label{sect:results}

\begin{figure}
\begin{center}
\includegraphics[width=7cm]{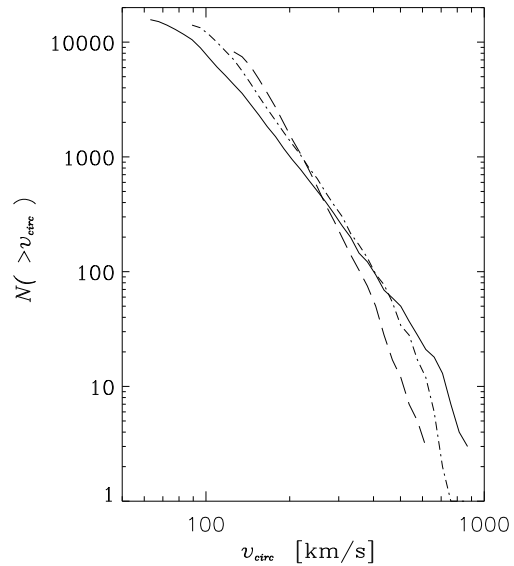}
\end{center}
\caption{Evolution of the cumulative number of halos with a circular
velocity larger than $v_{\rm circ}$. The solid, dot-dashed,
and dashed curves correspond to $z=0$, $z=1$, and $z=3$,
respectively. }
\label{fig:vel_z}
\end{figure}

For any study one needs to have a complete halo sample that is not
affected by the numerical details of the halo finding procedure. We
have tested the completeness of the halo samples using different
parameters for the halo finder.  For the given force and mass
resolution the halo samples do not depend on the numerical parameters
of the halo finder for halos with $v_{\rm circ} {_ >\atop{^\sim}} 100$
km/s {}\citep{gottloeber:halo}. In Fig.~\ref{fig:vel_z} we show the
cumulative number of halos with a circular velocity larger than a value
$v_{\rm circ}$, for redshifts $z=0$, $z=1$, and $z=3$,
respectively. Assuming a minimum circular velocity of 100 km/s the
samples are complete at $z \le 1$ but we are missing a small fraction
of halos with $v_{\rm circ} < 130$ km/s at $z=3$.

\subsection{Merging of halos}

According to hierarchical structure formation halos formed early and
grow during evolution due to accretion of matter and merging with
other halos. In particular merger events are important because they
are expected to lead to dramatic changes in the structure of dark
matter halos and the galaxies they harbor. In--falling objects may
damage or even destroy a stellar disk. The inflow of material may also
serve as a source of fresh gas and therefore increase the star
formation rate.  At the same time, collisions between halos may result
in shock heating of the gas, which would tend to delay or prevent star
formation for some period of time.

Following {}\citet{gottloeber:merging} we identify {\em major mergers}
as events when the mass of a halo grows by more than 25\% during a
time interval of about 0.5 Gyrs (approximate interval between
simulation outputs). In the above paper we showed
 that for redshifts $z < 2$ the
merger rate can be fitted by a simple power law $(1+z)^{3.0}$. This 
merger rate evolution is in very good general agreement with 
observations (e.g., \citet{lefevre:hubble-iv} measured a merger rate
varying with redshift as $\propto (1+z)^{3.2 \pm 0.6}$).
In addition, we found that evolution of the merger rate depends on the environment of the
halo: halos that end up in clusters and groups by $z=0$ have 
a steeper evolution of merger rate and 
a higher rate of major mergers at early epochs compared to isolated
``field'' halos. This is because clusters and groups form in the regions
that are overdense on large scales in which halos form and undergo 
the phase of active merging earlier than the overall field halo 
population. 

For the $z=0$ sample of halos with $v_{\rm circ}>100$~km/s, about 32~\% of
halos had one major merger in the past and additional 19~\% of halos
had two or more major mergers. Now let us consider the distribution of
epochs of the last major merger (relevant, for instance, for estimating
a fraction of halos that could host old disks such as that of the
Galaxy). We found that 55~\% of the halos located in clusters at $z=0$
underwent a major merger after redshift $z=4$, but that corresponding
fraction for the isolated halos is 43~\%. In contrast, the fraction of
isolated halos which underwent a major merger at a redshift $z<1$ is
somewhat higher (19~\%) than the corresponding fraction of halos in
clusters (14~\%); for $z<0.4$ (i.e., within the last $\approx 5$ Gyrs)
these fractions are 8~\% and 3.5~\%, respectively.  This reflects the
fact that due to the high internal velocity dispersion of halos in
clusters mergers are almost impossible. Since the merger rate of group
halos is high compared to the overall merger rate of halos at all
analyzed redshifts, the fraction of present-day group halos which
merged after a fixed redshift is always higher than the fraction of
isolated or cluster halos which merged after the same redshift.

Finally, let us consider one more effect. Due to the tidal
interactions halos in dense regions (i.e., virialized regions of
groups and clusters) tend to loose mass via tidal stripping. In order
to take this effect into account, we follow the mass evolution of all
halos and identify halos that lost more than 30\% of their maximum
(over their evolution) mass from $z=1$ to the present epoch.  One
would expect that galaxies hosted by such halos also lost their supply
of fresh gas so that no star formation was possible in the recent
past.

\subsection{Spatial distribution and evolution history}

In the preceding section we considered the overall fractions of halos
in different environments and with certain merger history classes.
This relatively straightforward analysis reveals existence of some
environmental dependency of halo evolution histories.  The goal of
this section is to carry out a more quantitative analysis of how
spatial clustering of different halo subsamples depends on evolution
histories of their halos. As discussed in the previous subsection,
major mergers (and tidal stripping) can be expected to result in
dramatic changes in the properties of galaxies (i.e., morphology and
color). One can expect, therefore, that the spatial distribution of
halos that experienced a recent merger or stripping event is different
from the distribution of the overall halo population. For example,
{}\citet{knebe:formation} found that massive halos undergoing mergers
at present exhibit a much stronger bias with respect to the dark
matter than relaxed halos do.
\begin{figure}
\begin{center}
\includegraphics[width=7cm]{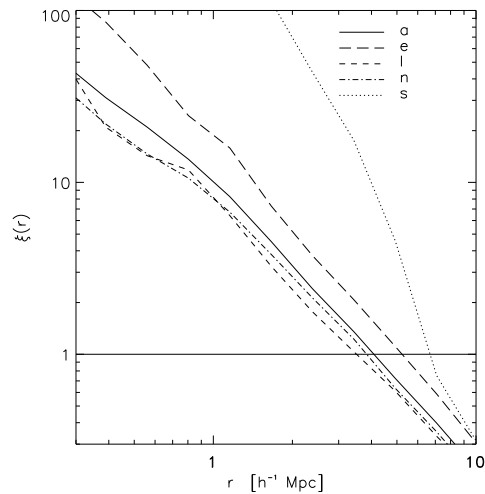}
\end{center}
\caption{The two--point correlation function of all (a, {\em solid} line) and subsamples
  of halos with different mass evolution histories (e: last major
  merger at $z>1$, {\em long-dashed} line; l: last merger at $z<1$,
  {\em short-dashed} line; n: never merged, {\em dot-dashed} line; and
  s: stripped halos, {\em dotted} line.}
\label{fig:corr}
\end{figure}

Figure~\ref{fig:corr} shows the two--point correlation function of all
halos (a) with $v_{\rm circ}>100$km/s (solid line) compared with that
of the subsamples of halos with different evolution histories.  We
divided the sample of all halos into four subsamples: halos which never
(n) underwent a major merger in the past, halos which underwent a major
merger before (e) and after (l) redshift $z=1$, and halos that lost
mass since $z=1$ (s). Note, that the stripped halos constitute a
separate sample, however every stripped halo belongs also per
definition to one of the other subsample. In particular, a substantial
part of ``stripped'' halos in clusters underwent a major merger before
redshift $z=1$, i.e. they belong to the sample (e) of halos.  They
lost most of their mass later on due to interactions.

Figure~\ref{fig:corr} shows clearly that the subsample of halos which
underwent the last major merger before redshift $z=1$ is more
clustered than the sample of all halos. This is not surprising since
most of such halos formed early in the regions of large-scale
overdensity and ended up in groups and clusters by the present epoch.
The stripped halos are even more biased with respect to the overall
halo population. This is also due to the fact that halos
that loose mass via stripping are located in the high-density regions
where tidal forces are strong.

\begin{figure}
\begin{center}
\includegraphics[width=7cm]{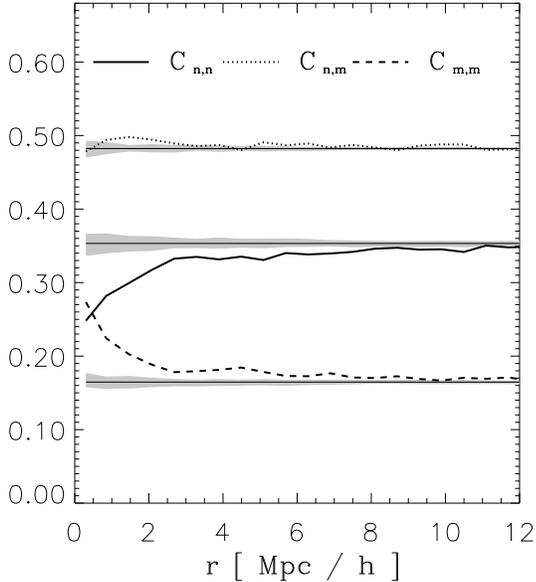}
\end{center}
\caption{Conditional cross--correlation function of halos with (m) 
and without (n) major merger in their evolution history.
The shaded area is obtained by randomizing the
assignment among the  halos (see text for details). The 
subsamples were selected from the total sample of halos with $v_{\rm
circ}>100$km/s.
\label{fig:cc2-merger}}
\end{figure}

Let us now consider the spatial distribution of different halos using the
mark correlation functions introduced in Sect.~\ref{sect:markcorr}.
To this end, we first split the total halo sample (i.e., sample of all
halos with circular velocities $v_{\rm circ}>100$km/s) into two
subsamples consisting of halos which experienced a major merger
(sample m) and halos which never experienced a major merger (sample
n), respectively. Figure~\ref{fig:cc2-merger} shows the conditional
cross--correlation functions (eq.~\ref{eq:cond-crosscorr}) of these
samples at $z=0$.  Positive $C_{m,m}$ on scales below 3\hMpc\ 
indicates that halos that experienced a major merger in their
formation history are relatively overabundant in close pairs of halos,
while halos without a major merger are underabundant.  No significant
cross--correlation between m and n exists.  Qualitatively, this
feature is independent from a lower cut in the circular velocity
$v_{\rm circ}$, but the amplitude of the effect is reduced if we
consider only the more massive halos with $v_{\rm circ}>150$km/s.
To investigate the significance of these deviations we performed a
non--parametric Monte Carlo test {}\citep{besag:simple}, similar to
the one used in {}\citet{kerscher:reflex} and \citet{kerscher:regular}.
Our null hypothesis is ``no mark correlation''. We simulate this null
hypotheses by keeping the positions of the halos fixed and 
randomly re-assigning the marks (in this case the class assignments) to
the halos. We repeat this to generate 99 realizations of this null
hypothesis. The shaded areas in Fig.~\ref{fig:cc2-merger} are the
one--$\sigma$ regions numerically determined from these samples.
To quantify the significance we have to define a distance
measure. Using $M=4$ equidistant radii $r_i$ in the range from
0.8\hMpc\ to 2.8\hMpc\ we define the ``distance'' of the $k$--th
sample to the expected value for no mark correlation:
\begin{equation}
d^k = \frac{1}{M}\sum_{i=1}^M 
\Big( C_{m,m}(r_i)-\frac{\varrho_m^2}{\varrho^2} \Big)^2 ,
\end{equation}
with the overall number density $\varrho$ and the number density
$\varrho_m$ of merged halos. Similarly we determine the distance
$d^{\rm halo}$ of the original halo sample to the null
hypothesis. Then we sort all the $d$'s and determine the position of
$d^{\rm halo}$. In this case $d^{\rm halo}$ is the fifth largest
distance, and we conclude that the original halo sample is
incompatible with the null hypothesis  "no mark correlation" at a
significance level of 95\% (see the comments by
{}\citealt{mariott:barnard} concerning the significance level).

\begin{figure}
\begin{center}
\includegraphics[width=7cm]{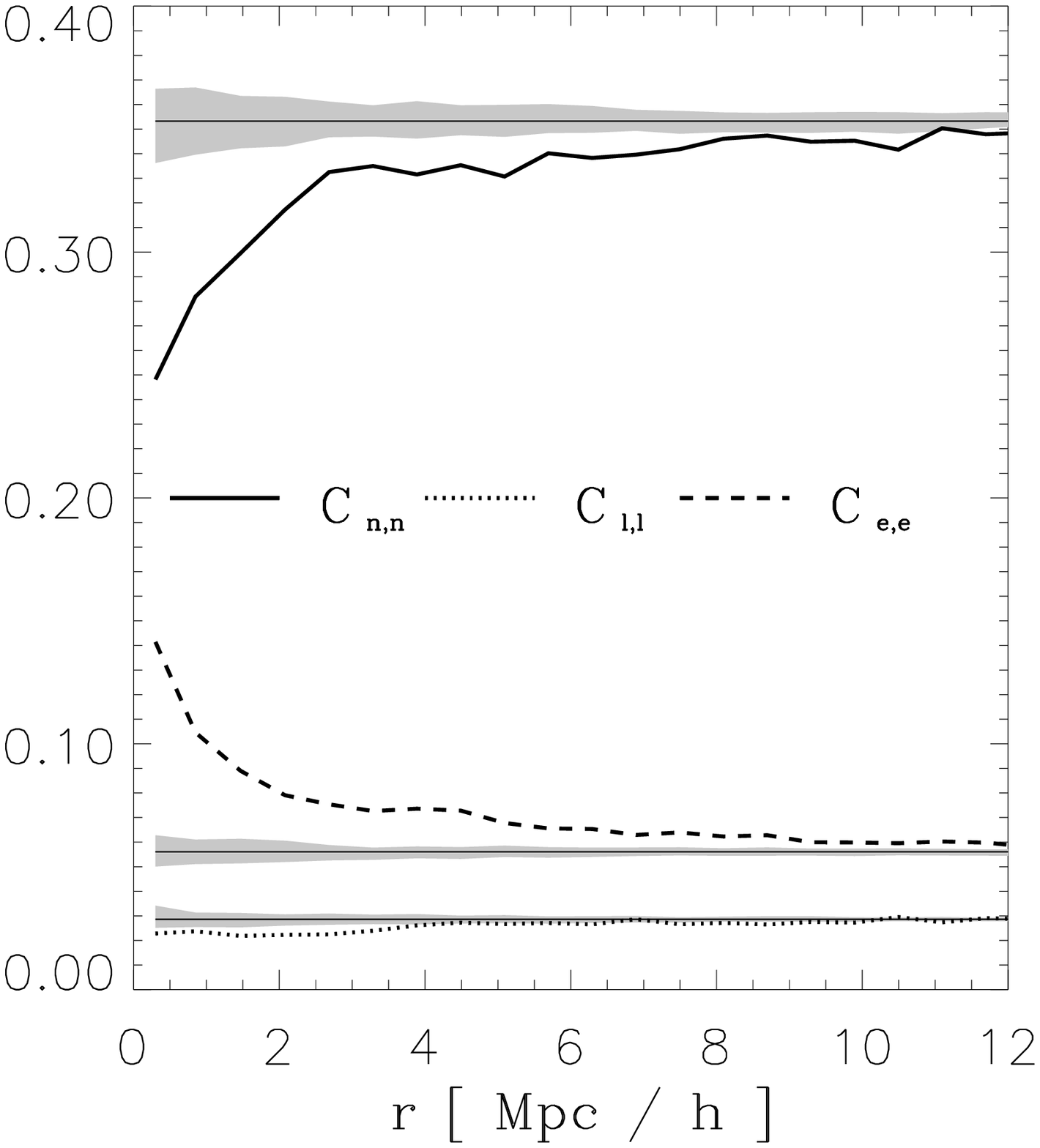}
\includegraphics[width=7cm]{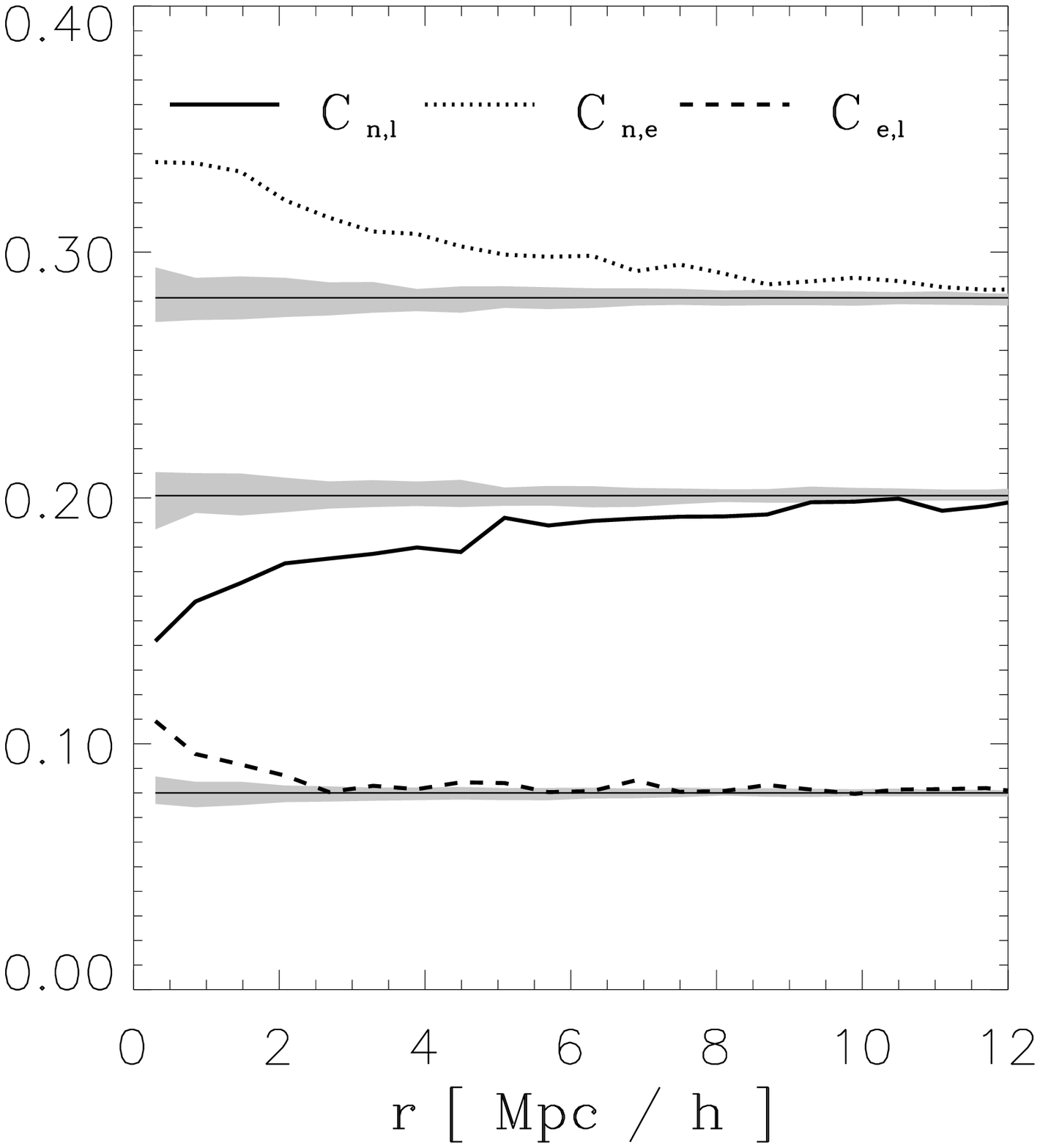}
\end{center}
\caption{ Conditional cross--correlations of halo subsamples 
  defined using the epoch or absence of the last major merger (e: the
  last major merger at $z>1$, l: at $z<1$, n: never occured). The
  shaded area is obtained by randomizing the assignment among the
  halos (see text for details). The subsamples were selected from the
  total sample of halos with $v_{\rm circ}>100$km/s.
\label{fig:cc3-merger-vc100}}
\end{figure}

Figure~\ref{fig:cc3-merger-vc100} shows the conditional
cross--correlations of the three subsamples which we considered above
using the two--point correlation function (Fig.~\ref{fig:corr}: 
the halos which never underwent a major
merger in the past, sample n; early major merger at $z>1$, sample
e; and late major merger at $0\le z\le1$, sample l).

As before, positive $C_{\rm e,e}$ at small separations indicates that for
objects at distances less than about 2 \hMpc\ the halos with a major
merger in their early formation history are relative over--abundant,
at the expense of halos without a major merger, as deduced from the
lowered $C_{\rm n,n}$. This signal is most prominent on scales below
3\hMpc\ but it extends out to 10\hMpc\ in agreement with the enhanced
two-point correlation function of that subsample
(Fig.~\ref{fig:corr}). We interpret this as indication of a high
number of early merged halos in clusters.  The signal has high
significance and it is not influenced by uncertainties in the
normalization of the correlation function that may be caused by
selection effects. Interestingly, the halos with a late major merger
show no excess correlations but rather a lowered abundance on small
scales, also manifested as the lower correlation function amplitude of
that subsample.

The lower panel of Fig.~\ref{fig:cc3-merger-vc100} shows the
cross-correlation of halos from different evolution classes. The
over-abundance of pairs of never and early merged halos reflects the
continuous accretion process onto high density regions. Infalling
isolated halos from less dense regions accrete onto higher density
regions with high velocity dispersions and, thus, low probability of
merging. Therefore, type-n halos can survive in the high-density
regions relatively long which  explains the increasing of
$C_{\rm n,e}$ towards small scales.  The opposite is true for never $n$
and late $l$ merged halos. Type-l halos located predominantly in
groups where mergers are more likely due to the lower velocity
dispersions.  The probability of accreting type-n halo (located close
to an type-l halo) to experience a merger in high-density regions is
therefore high. Many such halos will thus disappear (will become
l-halos) resulting in suppression of $C_{\rm n,l}$ amplitude at small
separations.

Note, that these features are qualitatively independent from a lower
cut in the circular velocity $v_{\rm circ}$, but the amplitude and 
the spatial range is reduced if we consider more massive halos with
$v_{\rm circ}>120$km/s or $v_{\rm circ}>150$km/s. This is due to the 
higher number of mergers within the low circular velocity halos.

\begin{figure}
\begin{center}
\includegraphics[width=7cm]{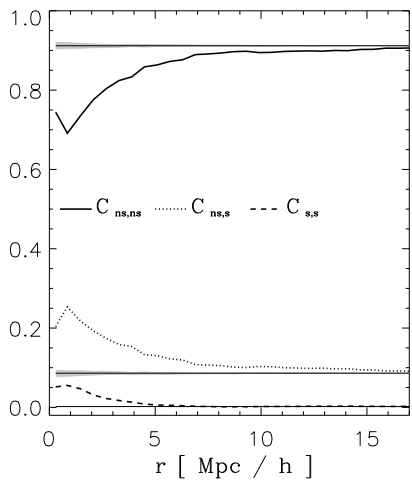}
\end{center}
\caption{ Conditional cross--correlations of halos 
  with (s for stripped) and without (ns) loss of mass in their mass
  evolution history (selected from the sample of halos with $v_{\rm
    circ}>100$km/s).
\label{fig:cc2-strip}}
\end{figure}

Let us now consider halos which lost a substantial part of their mass
due to tidal interactions. Figure~\ref{fig:cc2-strip} shows the
conditional cross--correlation function of the halos using two classes:
no stripping (sample: ns) and stripped (sample: s).  Positive $C_{\rm s,s}$
at $r<5h^{-1}{\rm\ Mpc}$ indicates that the number of stripped pairs
is strongly enhanced at these separations, whereas the number of
non--stripped pairs is reduced. This result is in accordance with the
strongly enhanced correlation function shown in Fig.~\ref{fig:corr}. In
fact, we expect to find stripped halos only in the environment of
clusters. The results for samples with a higher cut in the circular
velocity $v_{\rm circ}>150$km/s are very similar.

\subsection{Spatial distribution and circular velocity}

\begin{figure}
\begin{center}
\includegraphics[width=7cm]{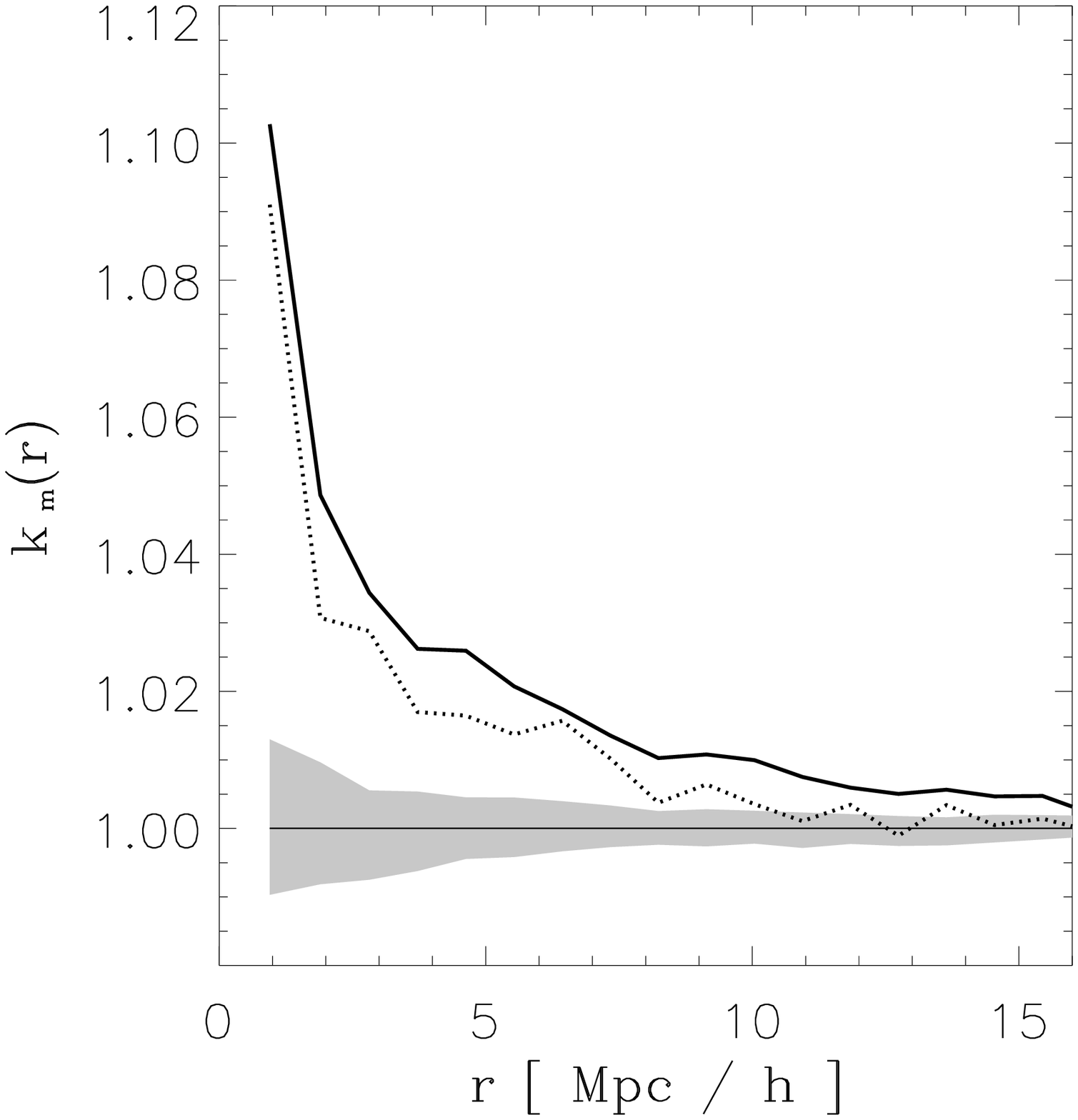}
\includegraphics[width=7cm]{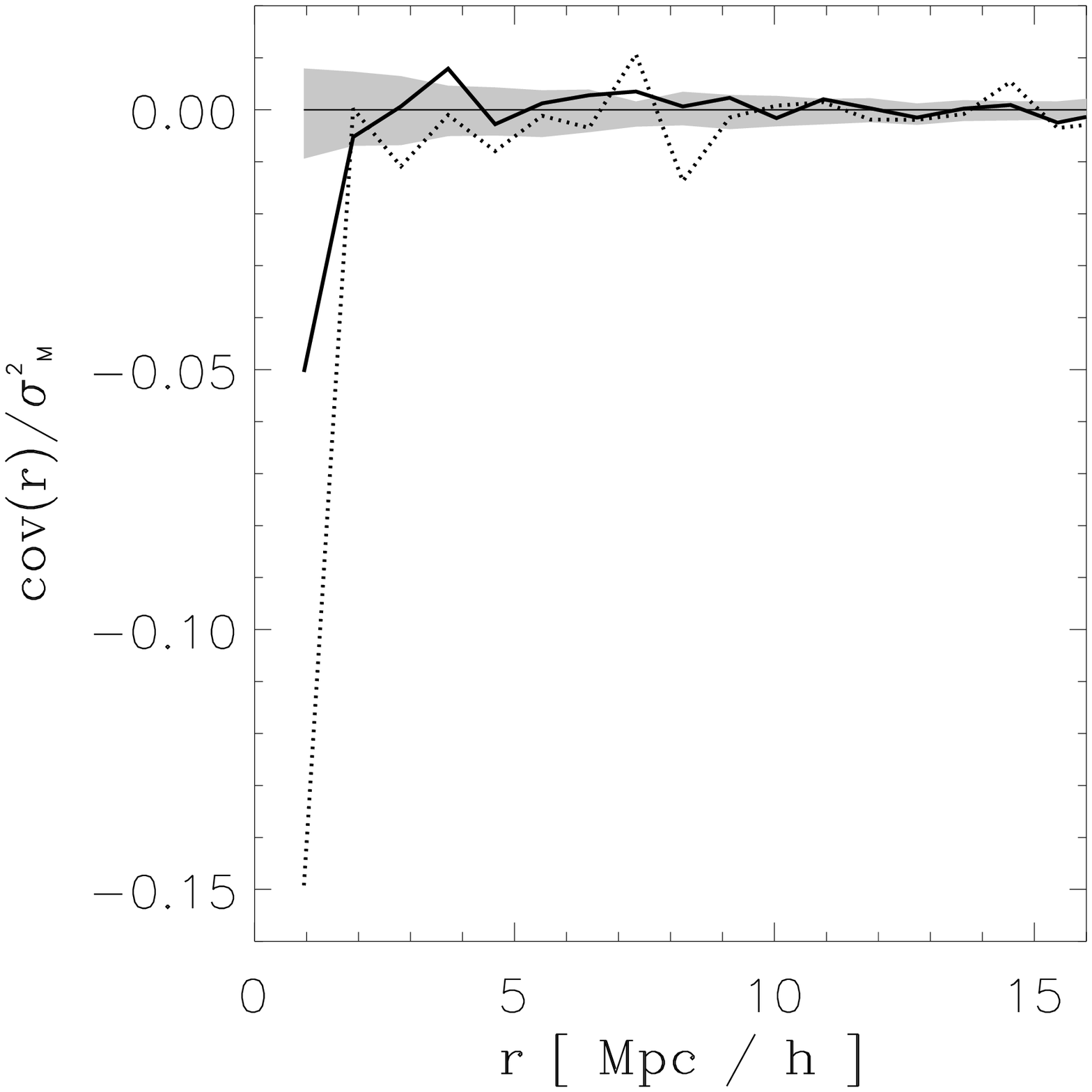}
\end{center}
\caption{ The mark correlation functions $k_m(r)$ and $\cov(r)$ (see 
Eqs.~\ref{eq:k_m} and \ref{eq:def-cov}) for
halos with the circular velocity $v_{\rm circ}$ as a scalar mark. Two
lower cut-offs have been used: $v_{\rm circ}>100$km/s (solid line)
and $v_{\rm circ}>150$km/s (dotted line).  The shaded area is
obtained by randomizing the mark among the halos.
\label{fig:mc-z0-vcirc}}
\end{figure}

As mentioned in Sec.~\ref{sect:numsim}, the mass and the maximum
circular velocity of halos are tightly related. At the same time, the
circular velocity can be determined more reliably in simulation as
well as in observations, either through direct measurement using
emission line width or rotation curve or via galaxy luminosity using
the Tully-Fisher and the Faber-Jackson relations.  Therefore, it is
interesting to explore galaxy mark correlations with the maximum
circular velocity as mark. This would mimic to some degree luminosity
segregation effects in observed galaxy samples.
Figure~\ref{fig:mc-z0-vcirc} shows the mark correlation functions of
halos at $z=0$ with the circular velocity as mark.  There is a strong
signal in $k_m(r)$ at small separations but the signal is significant
even out to 10\hMpc\ . This indicates that the mean circular velocity of
pairs of halos with separations below $\sim 10$\hMpc\ is larger than
the overall mean circular velocity $\overline{v}_{\rm circ}$ of the
parent halo sample. 
The negative signal of $\cov(r)$ is confined to small separations
($\lesssim 2$\hMpc). This is because pairs at small separations are
more frequently built from one halo with circular velocity larger than
$\overline{v}_{\rm circ}$ and the other halo with circular velocity
smaller than $\overline{v}_{\rm circ}$. Hence, this signal is
dominated by pairs of small-mass subhalos and massive parent halos.
The mark correlations results for both samples with $v_{\rm
  circ}>100$km/s and $v_{\rm circ}>150$km/s are shown. The mean mark
correlation function, $k_m(r)$, exhibits a slightly stronger signal
for the sample with a lower cut in circular velocity, but the signal
$\cov(r)$ is weaker for $v_{\rm circ}>100$km/s. The latter is due to
the considerably larger number of isolated small-mass halos; i.e. in
addition to pairs between large- and small-mass halos, for the $v_{\rm
  circ}>100$km/s sample there are many more small--small mass pairs.

Figure~\ref{fig:mc-zevolve-vcirc} shows the evolution of the
conditional covariance $\cov(r)$ of the circular velocity with
redshift.  The conditional covariance is negative at low redshifts out
to scales of 2\hMpc\, as discussed above.  At high redshifts ($z=2$ and
$z=3$), significant positive amplitude of $\cov(r)$ indicates that
pairs with similar circular velocities are overabundant. At these
redshifts the signal is significant out to the scale of $\sim 5${\hMpc\
 due to the large number of smaller-mass progenitors of the present
day halos.

\begin{figure}
\begin{center}
\includegraphics[width=7cm]{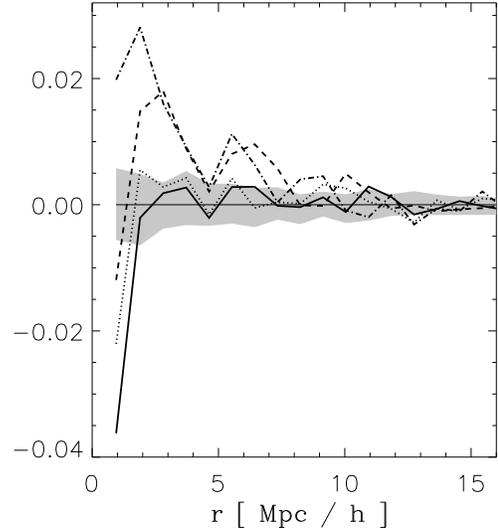}
\end{center}
\caption{ The redshift evolution of the mark correlation function
$\cov(r)$ for halos with the circular velocity $v_{\rm circ}$ as a
scalar mark and a lower cut-off of $v_{\rm circ}>100$km/s: $z=0$ (solid
line), $z=1$ (dotted line), $z=2$ (dashed line), $z=3$ (dashed dotted
line).  The shaded area is obtained by randomizing the mark among the
halos.
\label{fig:mc-zevolve-vcirc}
}
\end{figure}

\section{Discussion and conclusions}
\label{sect:conclusions}

In the previous section we used a novel statistical tool, the mark
correlation functions, to study clustering of galactic halos as a
function of their properties and environment in a high-resolution
numerical simulation of the $\Lambda$CDM cosmology.  We applied MCFs
using several types of continuous and discrete marks: maximum circular
velocity, $v_{\rm circ}$, of halos (continuous), merger mark
indicating whether halos experienced (m) or not (n) a major merger in
their evolution history, a stripping mark (s) indicating whether the
halo underwent a tidal stripping (i.e., mass loss) during its
evolution (discrete marks).  The halos which underwent major merger
(m) are further classified by the epoch of the merger: late (l; $z<1$)
and early (e; $z>1$) mergers.  Our main results are as follows.

The two-point correlation amplitude is different for the halo
subsamples with different marks. The halos that experienced an early
major merger or mass loss (e and s) are clustered considerably more
strongly than the overall halo population, while halos with late or no
mergers have correlation function amplitude below that of the overall
halo sample. This result indicates that halo clustering depends
sensitively on the details of their evolution history. If existence of
a major merger during halo evolution is related to the morphology of
galaxies that halos host, the above result indicates that early type
galaxies and galaxies in clusters and groups (hosted by halos that
undergo tidal stripping) should be clustered more strongly than the
late type galaxies and the overall galaxy population. Qualitatively,
such trend exists in the observed galaxy samples
\citep[e.g.,][]{hermit_etal96,guzzo_etal97,zehavi_etal02} implying that the
morphology-dependent clustering may be largely due to the overall
merger history of the galactic halos.

Using maximum circular velocity of halos as a continuous mark, we
found that at $z=0$ the mean circular velocity of pairs of halos with
separations $\lesssim 10$\hMpc\ is larger than the overall mean
circular velocity $\overline{v}_{\rm circ}$ of the parent halo sample
(manifested as significant enhancement of the mean mark at these
separations; see Fig~\ref{fig:mc-z0-vcirc}).  Moreover, the negative
mark covariance (Eq.~\ref{eq:def-cov}) at small separations shows an
enhanced abundance of pairs with halo circular velocities above {\em
  and} below the average circular velocity.  This mean circular
velocity enhancement increases steadily during the evolution of halos
from $z=3$ to $z=0$. The mark covariance, $cov(r)$, has a more
complicated behavior: it is negative at present, disappears at
redshift $z\sim 1$ and becomes positive at higher redshifts due to the
larger number of low circular velocity halos (the circular velocity
function of halos steepens at high redshifts, see
Fig.~\ref{fig:vel_z}).  Although the relation is not direct, the
maximum circular velocity of halos should correlate well with the
maximum circular velocity or velocity dispersion of the galaxies they
host. The enhanced mean circular velocity in small-separation pairs
should therefore correspond to the luminosity segregation or
luminosity-dependent clustering in the observed galaxies. The
luminosity dependence of galaxy clustering was recently convincingly
detected in both 2dF \citep{norberg_etal01} and SDSS
\citep{zehavi_etal02} galaxy surveys.

The mark correlation analysis indicate that galaxy-size halos ($v_{\rm
  circ}>100$km/s) which experienced a major merger in their evolution
history are over-abundant in pairs with separations $\lesssim 3$
\hMpc\ with respect to the overall halo population, while halos which
never experienced a major merger are under-abundant at these
separations. We find no significant cross-correlation between these
two halos classes. The overabundance of merger halos is due largely to
the halos which experienced a major merger relatively early ($z>1$); halos
with late ($z<1$) major merger are not over-abundant (this is also
manifested in the low amplitude of their two-point correlation
function relative to that of the overall halo population; see
Figs.~\ref{fig:corr}, \ref{fig:cc3-merger-vc100}). This result can be
interpreted as correlation between the time since the last major
merger and present-day environment of the halo (i.e., halos which
underwent an early major merger tend to be located in clusters and
groups).  The significance of the results was estimated by a
non--parametric Monte Carlo test which showed that the segregation and
anti-segregation have significance of $>95$\% in the distance range
0.8\hMpc\ to 2.8\hMpc\ .  Similarly, the probability of finding
  stripped halos in pairs of separations $\lesssim 5$\hMpc\ is twice
  higher than the corresponding probability for the overall halo
  sample. Halos which experienced early major mergers and/or mass loss
  due to tidal stripping are likely to host early type galaxies. In
  this case, the above mark correlation results for DM halos indicate
  that morphological segregation of galaxies may be due to the
  specifics of the mass evolution histories and environment of their
  parent halos.

The analysis presented in this paper showed that MCFs provide powerful,
yet algorithmically simple, quantitative measures of segregation in
halo spatial distribution with respect to their properties (e.g.,
maximum circular velocity) and merger history (e.g., time since the
last major merger). The mark correlation functions allow us to quantify
the degree of segregation as a function of scale and can be used to
quantify the differences in the spatial distributions of various galaxy
samples (similarly to the usual two-point correlation function) and, at
the same time, to study the interplay between the spatial clustering
and the distribution of galaxy properties (marks).  In this respect,
the MCFs are a natural extension of the spatial correlation functions
for studies where it is advantageous to consider mark and spatial
distributions simultaneously. We believe that this will make the mark
correlation functions very useful for analysis of spatial clustering
and segregation as a function of various galaxy properties in current
(SDSS and 2dF) and future (e.g., DEEP2) large redshift surveys.

\begin{acknowledgements}

S.G. acknowledges support from Deutsche Akademie der Naturforscher
Leopoldina with means of the Bundesministerium f\"ur Bildung und
Forschung grant LPD 1996. M.K. would like to thank the people at the AIP
for their hospitality on several occasions.  M.K. was supported by the
Sonder\-for\-schungs\-bereich 375-95 f\"ur Astro--Teilchen\-physik der
Deutschen For\-schungs\-ge\-mein\-schaft.  A.V.K. was partially 
supported by NASA through Hubble Fellowship grant from the Space
Telescope Science Institute, which is operated by the Association of
Universities for Research in Astronomy, Inc., under NASA contract
NAS5-26555.
\end{acknowledgements}

\end{document}